\documentclass[prd,aps,nofootinbib,superscriptaddress]{revtex4}
\usepackage{epsfig}
\usepackage{amsmath, slashed}
\usepackage{xcolor}
\usepackage{appendix}
\usepackage{dsfont}

\usepackage[normalem]{ulem}

\usepackage{bm}
\usepackage{amsfonts}

\allowdisplaybreaks
\begin{document}

\title{Kinematical higher-twist corrections in $\gamma^*  \to M_1  M_2 \gamma$ \\ I. Neutral meson production}

\author{Bernard Pire}
\email[]{bernard.pire@polytechnique.edu}
\affiliation{CPHT, CNRS, \'Ecole polytechnique, Institut Polytechnique de Paris, 91128 Palaiseau, France}
\author{Qin-Tao Song}
\email[]{songqintao@zzu.edu.cn}
\affiliation{School of Physics and Microelectronics, Zhengzhou University, Zhengzhou, Henan 450001, China}

\date{\today}

\begin{abstract}
{We carry out the  calculation of   kinematical higher-twist  corrections to the cross section of $\gamma^*  \to M_1 M_2 \gamma$ up to twist 4, where $M_i$ is a scalar or pseudoscalar neutral meson. 
The three independant helicity amplitudes are presented in terms of the twist-2 generalized distribution amplitudes (GDAs), which are 
important non-perturbative quantities for understanding the 3D structure of hadrons.
Since this process can be measured by BESIII in  $e^+ e^-$ collisions, we  perform the numerical estimate of the kinematical higher-twist  corrections 
by using the kinematics of BESIII.
We adopt the $\pi \pi$ GDA extracted from Belle measurements and the asymptotic $\pi \pi$ GDA
to study the size of the kinematical corrections in the case of pion meson pair, 
and a model $\eta \eta$ GDA is  used to see the impact  of target mass corrections $\mathcal O(m^2/s)$ for $\gamma^*  \to \eta \eta \gamma$. 
Our results show that the kinematical higher-twist  corrections account for
$\sim 20\%$ of the cross sections at BESIII  on the average, and  it is necessary to  include them if one tries to extract GDAs from experimental measurements precisely.
We also comment the case of $\pi^0 \eta$ production which is important for the search of hybrid mesons.

  }

\end{abstract}

\maketitle

\date{}

\section{Introduction}
\label{introduction}
Generalized distribution amplitudes (GDAs) \cite{Muller:1994ses, Diehl:1998dk, Diehl:2000uv, Polyakov:1998ze,Pire:2002ut} are important non-perturbative functions which reveal the 3D partonic  structure of hadrons, and they correspond to the amplitudes of  the soft processes $q \bar{q} \to h \bar{h}$ or $gg \to h \bar{h}$.
GDAs were firstly investigated in  $\gamma^{\ast}  \gamma \to  h \bar{h}$ with large photon virtuality and small 
invariant mass of hadron pair  to satisfy the QCD factorization.  This process is known as the 
$s$-$t$ crossed channel
of deeply virtual Compton scattering (DVCS), from which 
generalized parton distributions (GPDs) \cite{Diehl:2003ny,Belitsky:2005qn, Boffi:2007yc, Goeke:2001tz} are probed.
Recently, measurements of $\gamma^{\ast}  \gamma \to  h \bar{h}$  were  released for neutral pion pair \cite{Belle:2015oin} and neutral kaon pair \cite{Belle:2017xsz} production by the Belle collaboration at KEK,
and the $\pi \pi$ quark GDA was extracted from the cross section of $\gamma^{\ast}  \gamma \to  \pi^0 \pi^0$ \cite{Kumano:2017lhr}.
In addition to  $\gamma^{\ast}  \gamma \to  h \bar{h}$, GDAs can also be  accessed in  the crossed reaction \cite{Lu:2006ut}: 
\begin{equation}
  \gamma^{\ast}(q_1)   \to  h (p_1) \bar{h}(p_2) \gamma (q_2) \,,
  \label{reaction}
\end{equation}
which may be studied in the electron-positron annihilation process
\begin{equation}
  e^-(k_1) e^+(k_2)   \to  h (p_1) \bar{h}(p_2) \gamma (q_2) \,,
  \label{fullreaction}
\end{equation}
where the large scale $Q^2=q_1^2$ is now timelike; a first access to  this process was released  by BaBar \cite{BaBar:2015onb} in the charged meson channel case, and  future results should be obtained at BESIII and Belle (Belle II).

There exists a basic difference between the neutral (say $\pi^0 \pi^0$) production channel and the charged one (say $\pi^+ \pi^-$), due to the charge conjugation property of the $\pi \pi$ state. Since the $\pi^+ \pi^-$ pair can be produced both with $C=+$ and $C=-$ charge conjugation, the QCD amplitude (\ref{reaction}) can interfere with the QED process, known as Initial State Radiation (ISR) :
\begin{equation}
    e^-(k_1) +  e^+(k_2) \to \gamma^*(q'_1) + \gamma (q_2) ~~; ~~ \gamma^*(q'_1) \to h (p_1) \bar{h}(p_2) \,,
    \label{QED}
\end{equation}
where the $h (p_1) \bar{h}(p_2)$ pair is produced in a $C=-$ state. The amplitude of the ISR process (\ref{QED}) does not depend on GDAs, and is readily calculated with the help of the measured $\pi$ meson timelike electromagnetic form factor. The ISR process turns out to dominate\footnote{The typical parameter which controls the magnitude of the relative contributions of both processes to the cross-section is $s/W^2$.} the QCD process in most kinematics, which renders inefficient the extraction of GDAs from integrated $\pi^+ \pi^-$ cross-sections, but allows us to extract the QCD contribution - and hence the GDAs - at the amplitude level through cleverly defined asymmetries, taking advantage of the different $C$ parities of the meson pair selected by the two processes. This is quite reminiscent of the usual procedure in DVCS or TCS measurements where the interference of the QCD process with the Bethe-Heitler contribution populates interesting asymmetries.
We shall address these questions in a forthcoming study and restrict here to the neutral pseudoscalar meson pair case, namely $\pi^0 \pi^0$, $\eta \eta$ and $\pi^0 \eta$ channels, where the process (\ref{QED}) does not contribute.

 On the one hand, GDAs are important inputs for the three-body  decays of $B$ mesons, which are used to study  the Cabibbo-Kobayashi-Maskawa (CKM) matrix \cite{Chen:2002th,Wang:2015uea, Li:2016tpn, Jia:2021uhi}. On the other hand,
 GDAs and GPDs are key objects to investigate the matrix elements of the energy-momentum tensor (EMT) for hadrons, which are expressed in terms of the EMT form factors.
 In principle, one can not measure  the EMT form factors of hadrons directly by experiment since  the gravitational interactions between hadrons and gravitons are too tiny to probe. 
However, GDAs and GPDs can be accessed via electromagnetic interactions,
as a consequence,  the studies of  GDAs and GPDs are quite valuable.
Many important physical quantities of hadrons can be obtained through the study of the EMT form factors, e.g.,
 mass, pressure and shear force distributions of hadrons\cite{Polyakov:2002yz, Goeke:2007fp, Mai:2012cx, Polyakov:2018zvc, Lorce:2018egm, Burkert:2018bqq,  Kumericki:2019ddg,Dutrieux:2021nlz,Burkert:2023wzr}. Let us also note that the production of two different mesons, for example $\gamma^{\ast} \to \gamma   \pi \eta$ where the $\pi \eta$ GDA is accessed,   can be also used to search for the hybrid meson ($J^{PC}=1^{-+}$)\cite{Anikin:2006du} and to investigate the shear viscosity of quarks in hadronic matter \cite{Teryaev:2022pke}.

GPDs and GDAs can currently be accessed at many experimental facilities, but in a quite limited range of the hard scale $Q^2$ which is for instance the virtuality of the incoming photon. Compared with the leading-twist cross sections,  the higher-twist corrections are thus not negligible
considering the energy scales of present and near future experimental measurements. In order to extract GPDs and GDAs precisely, one needs to include the  higher-twist contributions to the cross sections. However, higher-twist GPDs and GDAs  are required to describe the  higher-twist corrections, and  this will make the analysis difficult when one tries to extract GPDs and GDAs from the experimental measurements. 
Recently,  a separation of kinematical and dynamical contributions in the operator product of two electromagnetic currents $T \{j_{\mu}^\text{em}(z_1x)j_\nu^\text{em} (z_2x) \}$  was proven in Refs.\,\cite{Braun:2011zr,Braun:2011dg,Braun:2011th, Braun:2022qly}, and  these operator results can be applied to the off-forward hard reactions
such as $\gamma^{\ast} h \to \gamma h$, $\gamma h \to \gamma^{\ast} h$, $\gamma^{\ast}  \gamma \to  h_1 h_2$ and $\gamma^{\ast}   \to  h_1 h_2 \gamma$, where GPDs and GDAs can be accessed.
If one  includes the kinematical higher-twist contributions to the leading-twist cross sections, then
only  the leading-twist GPDs and GDAs are involved.
 This does not prevent genuine higher twist contributions from being potentially important; progress in their studies is indeed much needed.
The kinematical higher-twist corrections can be considered as a generalization of the target mass corrections \cite{Nachtmann:1973mr}, which are applied to the reaction of Deep Inelastic Scattering (DIS), and such corrections were already included in Ref.\,\cite{Sato:2016tuz} where the parton distribution functions were extracted. 
However, the kinematical higher-twist corrections  are more complicated in the off-forward hard reactions
due to the  higher-twist operators which are reduced to the total derivatives of the leading-twist ones, and these operators do not contribute in DIS since their forward matrix elements  vanish.

The kinematical higher-twist corrections were given up to the twist-4 accuracy  for the DVCS amplitude with a (pseudo)scalar target \cite{Braun:2012bg} and a spin-1/2 target \cite{Braun:2012hq, Braun:2014sta}.
The theoretical results were applied to the recent DVCS measurements by the JLAB Hall A collaboration\cite{JeffersonLabHallA:2022pnx}.
The authors of Ref.\,\cite{Lorce:2022tiq,Lorce:2022cze} also estimated  the
 kinematical higher-twist corrections for $\gamma^*  \gamma \to M \bar M$ with a (pseudo)scalar meson pair.
 All these theoretical studies suggest that the kinematical higher-twist corrections are not negligible in realistic experiments;
 besides, experimental measurements of DVCS also indicate that the kinematical corrections are sizeable in the cross section  and have to be taken into account \cite{JeffersonLabHallA:2015dwe, Defurne:2017paw}. 
 In this work, we intend to calculate the kinematical higher-twist corrections in  $\gamma^*   \to M_1  M_2 \gamma$, and this process can be measured at BESIII in future. The kinematics of BESIII measurements on this process   will be similar to the Belle (Belle II) measurements on 
 $\gamma^*  \gamma \to M_1  M_2 $, whose cross sections were released recently \cite{Belle:2015oin,Belle:2017xsz}.
 In this case, the GDAs  can be extracted by combining 
 $\gamma^*   \to M_1 M_2 \gamma$ and  $\gamma^*  \gamma \to M_1  M_2 $. Moreover, one can study the universality of GDAs by comparing the two processes, taking into account of the fact that the virtual photon is timelike in the former and it is spacelike in the latter.

In Sec.\,\ref{kinematics}, we discuss the kinematics of $\gamma^*   \to M \bar M \gamma$, and the cross section is presented in terms of helicity amplitudes.  
We carry out a  complete calculation of kinematical higher-twist 
corrections to the helicity amplitudes up to twist 4 in Sec.\,\ref{fr},  and the numerical estimate of the kinematical higher-twist corrections are also presented.
Our results are summarized in Sec.\,\ref{summary}.

\section{Kinematics and helicity amplitudes of $\gamma^* \to M \bar{M}  \gamma$ }
\label{kinematics}

\begin{figure}[htp]
\centering
\includegraphics[width=0.55\textwidth]{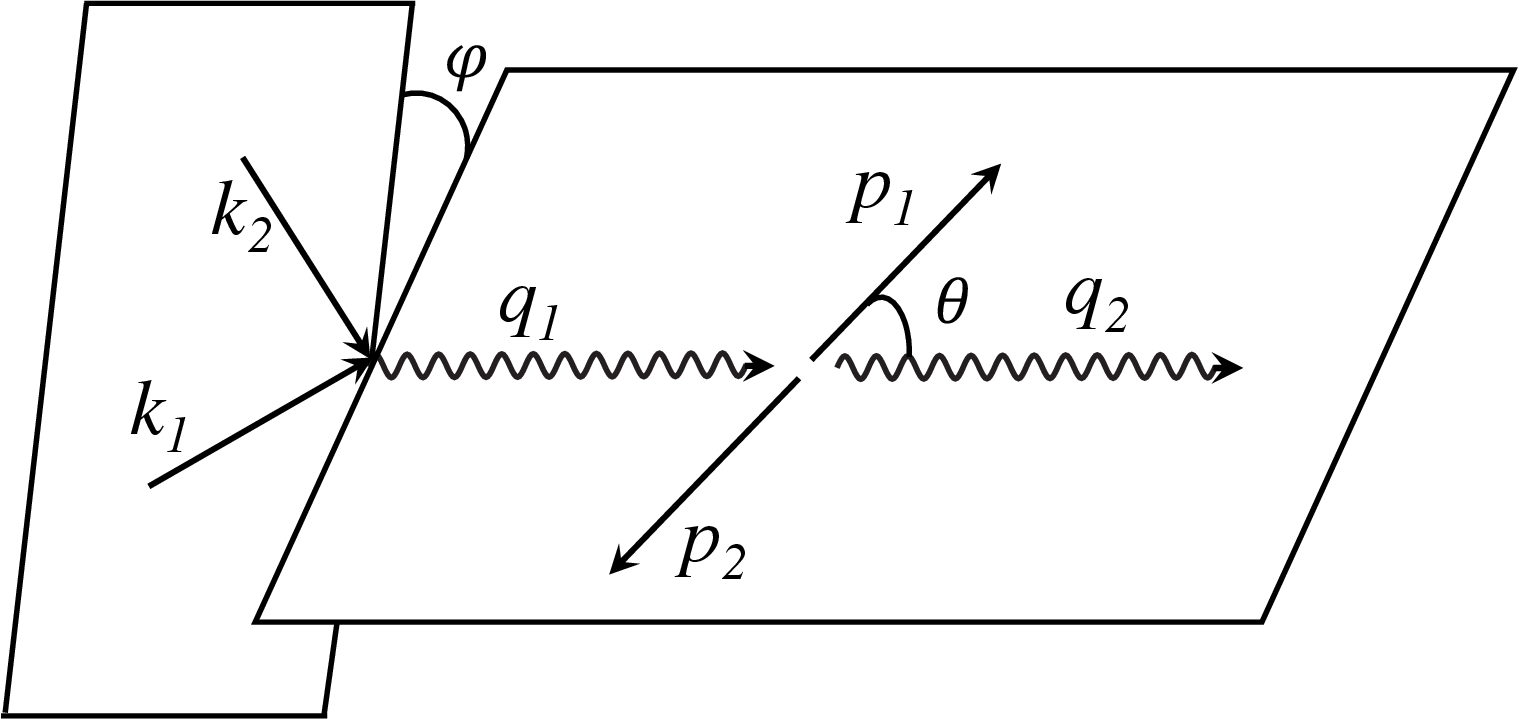}
\caption{Kinematics of  $e^-(k_1) e^+(k_2)  \to  \gamma^*(q_1) \to M(p_1) \bar{M}(p_2)  \gamma(q_2)$ in the center-of-mass frame of the meson pair, and the direction of the photons is chosen to the z axis.  }
\label{fig:kin}
\end{figure}
If the center-of-mass energy $\sqrt{s}$ is large enough to satisfy the QCD factorization  in the process   $e^- e^+ \to  \gamma^* \to M \bar{M}  \gamma$, then the amplitude can be factorized into a perturbative subprocess $\gamma^* \to q \bar{q}  \gamma$ and  
a two-meson GDA which describes the amplitude of $  q \bar{q} \to M \bar{M}$ \cite{Lu:2006ut}.
In Fig.\,\ref{fig:kin}, we show the polar angle $\theta$ and the azimuthal angle $\varphi$ in the  center-of-mass  frame of the meson pair,  for convenience we choose a coordinate system with the z axis along the direction of photons, so that the momenta of the mesons lie in the x-z plane. The polar angle $\theta$ can be expressed as
\begin{align}
\cos{\theta}=\frac{q_1\cdot(p_2-p_1)}{\beta_0\,(q_1\cdot q_2)}, \qquad \beta_0=\sqrt{1-\frac{4 m^2}{\hat{s}}}, 
\label{eqn:pola}
\end{align}
where $m$ is the meson mass and $\hat{s}=W^2=(q_1-q_2)^2=(p_1+p_2)^2$. Similarly, the azimuthal angle $\varphi$ is also given in terms of Lorentz invariants,
\begin{align}
\sin{\varphi}=\frac{   4  \epsilon_{\alpha \beta \gamma \delta}  q_1^{\alpha} q_2^{\beta}p_1^{\gamma} k_1^{\delta}  }{\beta_0 \sin{\theta} \sqrt{us \hat{s}(\hat{s}-u-s) }}
\label{eqn:amu}
\end{align}
with $ \epsilon_{0123}=1$, $s=(k_1+k_2)^2$ and $u=(k_1-q_2)^2$.
Two lightlike vectors $n$ and $\tilde{n}$ are chosen with the help of  the momenta of the timelike virtual photon $q_1$ and the  real photon $q_2$,
\begin{align}
\tilde{n} = q_1-(1+\tau)q_2, \qquad  n=q_2,
\label{eqn:lightlike}
\end{align}
where  $\tau=\hat{s}/(s-\hat{s})$. 
The momentum $\Delta=p_2-p_1$ can be written as  $\Delta=\zeta_0(\tilde{n}-\tau n) + \Delta_T$, and $\Delta_T$   is the transverse component.   $\zeta_0$ is a parameter which  is defined as
\begin{align}
\zeta_0= \frac{(p_2-p_1)\cdot n}{(p_2+p_1)\cdot n},
\label{eqn:skewness}
\end{align}
and it is related to the polar angle $\theta$ as $\zeta_0=\beta_0 \cos{\theta}$.
We can obtain $\Delta_T^2=4m^2-(1-\zeta_0^2)\hat{s}$  by the on-shell condition.

To describe the process  $\gamma^* \to M \bar{M}  \gamma$, one needs to define the amplitude
\begin{align}
A_{\mu \nu}=i\int d^4x\, e^{-ir\cdot x} \langle \bar{M}(p_2) M(p_1)  | \, 
T \{ j_{\mu}^{\text{em}}(z_1x)  j_\nu^{\text{em}} (z_2x) \} \, | 0 \rangle , \! \!
\label{eqn:amp0}
\end{align}
where $z_1$ and $z_2$ are real constants with the constraint $z_1-z_2=1$, and 
$r=z_1q_1-z_2 q_2$ is used.
This amplitude can be further written in terms of helicity amplitudes 
by considering the electromagnetic gauge invariance \cite{Braun:2012bg}
\begin{align}
A^{\mu \nu}=-A^{(0)}\,g_{\perp}^{\mu\nu}+A^{(1)}\, 
\left(\tilde n^\mu-(1+\tau)n^\mu\right) \frac{\Delta_{\alpha}g_{\perp}^{\alpha \nu}}{\sqrt{s}} 
+\frac{1}{2}\, A^{(2)}\, \Delta_{\alpha}\Delta_{\beta}(g_{\perp}^{\alpha \mu} g_{\perp}^{\beta \nu}- \epsilon_{\perp}^{\alpha \mu}   \epsilon_{\perp}^{\beta \nu} )+ A^{(3) \mu}\, n^{\nu},
\label{eqn:amp1}
\end{align}
and the transverse tensors $ g_{\perp}^{\mu \nu}$ and $\epsilon_{\perp}^{ \mu \nu}$  are defined as
\begin{align}
g_{\perp}^{\mu \nu} =g^{\mu\nu}-
\frac{n^{\mu}\tilde{n}^{\nu}+n^{\nu}\tilde{n}^{\mu}}{n\cdot \tilde{n}}, \qquad    
\epsilon_{\perp}^{\mu \nu} = \epsilon^{\mu \nu \alpha \beta}\,
 \frac{\tilde{n}_{\alpha} n_{\beta}}{n\cdot \tilde{n}}.
\label{eqn:gt}
\end{align}
The longitudinal  polarization vector $\epsilon_0$ exists in addition to the transverse ones
$\epsilon_{\pm}$ for the virtual photon.
In the center-of-mass frame of the meson pair as shown in Fig.\,\ref{fig:kin}, the polarization vectors read
\begin{align}
&\epsilon_{0}^{\mu}=\frac{1}{\sqrt{s}}(|q_1^3|, 0, 0, q_1^0), \quad
\epsilon_{\pm}^{\mu}=\frac{1}{\sqrt{2}}(0, \mp 1, -i, 0),
\label{eqn:pol-virt}
\end{align}
and the transverse polarization vectors $\tilde{\epsilon}_{\pm}$ of the real photon are the same  as the ones of the virtual photon. 
Then, the three independent helicity amplitudes are given by
\begin{align}
&A_{++}=A_{--}=A^{(0)}, \quad \quad 
A_{0 \pm }=-A^{(1)} (\Delta \cdot \epsilon_{\mp}),  \nonumber \\
&A_{\pm \mp}=-A^{(2)} (\Delta \cdot \epsilon_{\pm})^2.
\label{eqn:am-hel-ind}
\end{align}
where the notation $A_{i j}=  \epsilon_{i}^{\mu}   \tilde{\epsilon}_{j}^{\ast \nu } A_{\mu \nu}$ is used, and only
$A_{++}$ can receive twist-2 contribution at leading order of $\alpha_s$. 
In Ref.\,\cite{Lu:2006ut}, $A_{++}$ was given in terms of the $\pi \pi$ GDA at twist-2 for the process of $e^- e^+  \to  \gamma^*\to \pi \pi \gamma$,
and the twist-2  $\pi \pi$ GDA  is defined as \cite{Diehl:1998dk, Diehl:2000uv}
\begin{align}
\langle \bar{M}(p_2) M(p_1)  | \,\bar{q}(z_1 n) \slashed{n}q(z_2 n)\, | 0 \rangle  
=  2 P\cdot n \int dz \,e^{2i \left[ zz_1+(1-z)z_2 \right] P\cdot n}\, \Phi_q(z,\zeta_0, \hat{s}),
\label{eqn:gda}
\end{align}
where $z$ is  the momentum fraction of the quark,  $P$ is the average momentum of the meson pair $P=(p_1+p_2)/2$, and the real constants $z_1$ and $z_2$ do not have to satisfy $z_1-z_2=1$.
Note that the GDAs depend on a renormalization scale $\mu^2$ which one usually takes as $\mu^2=s$.

The differential cross section can be expressed in terms of the helicity amplitudes  for  $e^- e^+  \to  \gamma^*\to M \bar{M} \gamma$,
\begin{align}
\frac{d \sigma}{d\hat{s} \,du\, d(\cos \theta)\, d\varphi}=&
\frac{\alpha_{\text{em}}^3 \beta_0}{16 \pi s^3 } \,\frac{1}{1+\epsilon}   \,\Big[ |A_{++}|^2+ |A_{-+}|^2+2\epsilon\, |A_{0+}|^2  -2  \text {sgn}( \tau)\sqrt{\epsilon(1-\epsilon)}  \nonumber \\ 
 &\times   \text{Re}(A_{++}^{\ast} A_{0+} -A_{-+}^{\ast} A_{0+})  \cos \varphi   + 2 \epsilon \,  \text{Re}(A_{++}^{\ast} A_{-+}) \cos (2\varphi)  \Big],
\label{eqn:epho-cro}
\end{align}
where  $M \bar{M}$ is the  pseudoscalar meson pair with  even charge conjugation and 
 the parameter $\epsilon$ is defined as
\begin{align}
\epsilon=\frac{y-1}{1-y+\frac{y^2}{2}}, \qquad y=\frac{q_1\cdot q_2}{k_1\cdot q_2}.
\label{eqn:polar}
\end{align}
 $ \text {sgn}(\tau) =  |\tau|/\tau$ is the sign function with $\tau=\hat{s}-s-2u$.

\section{Results}
\label{fr}

\subsection{Theoretical amplitudes in terms of GDAs }
Recently, Braun \textit{et al.} have proved that 
the kinematical   contributions  can be separated from dynamical ones in the time-ordered product of two electromagnetic currents
$i\,T \{j_{\mu}^{\text{em}}(z_1x)j_{\nu}^{\text{em}} (z_2x) \}$ \cite{Braun:2011zr,Braun:2011dg,Braun:2011th}. 
If we introduce the kinematical higher-twist contributions to the twist-2 cross sections, 
one can improve the description of  reactions where two photons are involved  without any knowledge of the genuine higher-twist distributions.  This is very helpful when one intends to extract the twist-2 distributions from the measurements of the cross sections, since taking into account of genuine higher-twist distributions would imply  more parameters to be used in the analysis.
At twist-4 accuracy,  the kinematical contributions in $i\,T \{j_{\mu}^{\text{em}}(z_1x)j_{\nu}^{\text{em}} (z_2x) \}$ can be  written  as \cite{Braun:2011zr,Braun:2011dg,Braun:2011th}
\begin{align}
T_{\mu \nu}=\frac{-1}{\pi^2 x^4 z_{12}^3}
\left \{
x^{\alpha } \left[S_{\mu \alpha \nu \beta   } 
\mathbb{V}^{\beta}-i\epsilon_{\mu \alpha \nu \beta }\mathbb{W}^{\beta}
\right]  
+x^2 \left[ (x_{\mu} \partial_{\nu} + x_{\nu} \partial_{\mu}   )  \mathbb{X} + (x_{\mu} \partial_{\nu} - x_{\nu} \partial_{\mu}   )  \mathbb{Y} \right]
 \right \},
\label{eqn:kine-t4}
\end{align}
where the notation $z_{12}=z_1-z_2$ is used, and
the tensor  $S^{ \mu \alpha \nu \beta }$ is defined by
\begin{align}
S^{ \mu \alpha \nu \beta }=g^{\mu \alpha} g^{\nu \beta}-g^{\mu \nu}g^{\alpha \beta}+g^{\mu \beta} g^{\nu \alpha}.
\label{eqn:app2s}
\end{align}
$\mathbb{V}_{\mu}$  and $\mathbb{W}_{\mu}$  contain all twists starting from twist 2 to twist 4, whereas 
$\mathbb{X}$ and $\mathbb{Y}$ are purely twist 4 operators,  the detailed expressions of
 them can be found in Appendix A of our previous work \cite{Lorce:2022tiq}.

In this work we shall use Eq.\,(\ref{eqn:kine-t4}) to calculate the  kinematical 
higher-twist contributions in the reaction  $\gamma^*  \to M \bar{M}\gamma$, where $M \bar{M}$ is the  scalar meson pair with  even charge conjugation. The spinor formalism \cite{Braun:2008ia, Braun:2009vc} is used for $T_{\mu \nu}$ in the calculation, and it will help us to figure out the twist of the matrix elements for the operators much easier. Even though the final amplitudes are presented in terms of GDAs,  the double distributions  \cite{Teryaev:2001qm} are used to calculate the helicity amplitudes. 
Those techniques are explained in Appendix \ref{app-b},
and the helicity amplitudes are written as 
\begin{align}
A^{(0)}&= \chi \left\{  \left(1+ \frac{\hat{s}}{2s}\right) \int_0^1 dz\, \frac{\Phi(z, \eta, \hat{s})}{1-z} 
+\frac{\hat{s}}{s}\int_0^1  dz \, \frac{\Phi(z,\eta, \hat{s})}{z}\, \ln(1-z)
\right. \nonumber \\
&\qquad +\left. \left(\frac{2\hat{s}}{s} \,\eta   +\frac{\Delta_T^2}{\beta_0^2 s} \frac{\partial}{\partial \eta} \right)
\frac{\partial}{\partial \eta}
  \int_0^1 dz \,\frac{ \Phi(z,\eta, \hat{s}) }{z} \left[ \frac{\ln(1-z)}{2} +\text{Li}_2(1-z) -\text{Li}_2(1)  \right]
  \right \}, \nonumber \\
A^{(1)}&=  \frac{2\chi}{\beta_0 \sqrt{s}} \frac{\partial}{\partial  \eta }   \int_0^1 dz \, \Phi(z, \eta, \hat{s}) \,
\frac{\ln(1-z)}{z}, \nonumber \\
A^{(2)}&=\frac{2 \chi}{ \beta_0^2 s} \frac{\partial^2}{\partial \eta^2}       \int_0^1 dz  \,\Phi(z, \eta, \hat{s}) \,
\frac{2z-1}{z}\, \ln(1-z),
\label{eqn:int-gdas-a}
\end{align}
where $\eta=\cos \theta$,  $\chi=5e^2/18$,
and $\Phi=\Phi_u+\Phi_d$ for the isosinglet meson pair such as $\pi \pi$ and $\eta \eta$. 
In case of a charge conjugation-even $K$ meson pair, we just need to replace $\chi \Phi$ with 
$e_u^2 \Phi_u+e_d^2 \Phi_d +e_s^2 \Phi_s$ in the above amplitudes since 
the GDA for $s$ quark is also required.
We note that although Eq.\,(\ref{eqn:kine-t4}) is dependent on $z_1$ and $z_2$, this dependence is eliminated by the constraint $z_1-z_2=1$  in physical  amplitudes, which indicates that translation invariance is recovered.
The  existence of correction $\mathcal O(\hat{s}/s)$ is obvious in the amplitudes, while the correction $\mathcal O(m^2/s)$ exists in the term $\Delta_T^2/ s$ due to $\Delta_T^2=4m^2-(1-\zeta_0^2) \hat{s}$.

The general expressions of GDAs were firstly derived in \cite{Diehl:2000uv}, in this work we only need 
the charge conjugation-even ones, and they are given by
\begin{align}
\Phi(z, \cos \theta, \hat{s})=6\, z(1-z) \sum_{\substack{ n=1\\ n\,\text{odd} }}^{\infty} \sum_{\substack{l=0\\ l\,\text{even}}}^{n+1}
\tilde{B}_{nl}(\hat{s}) \,C_n^{(3/2)}(2z-1)\, P_l(\cos \theta),
\label{eqn:gda-expression}
\end{align}
where $C_n^{(3/2)}(x)$ are Gegenbauer polynomials,
and $P_l(x)$ are Legendre polynomials with $l$ indicates the angular momentum of the scalar meson pair.
The nonvanishing helicity-flip amplitudes $A_{-+}(A^{(2)})$ and $A_{0+}(A^{(1)})$ indicate the existence of the D-wave or higher-wave GDA. In Eq.\,(\ref{eqn:int-gdas-a}), it seems that  the amplitudes are divergent  if $ z\rightarrow 0$ or $ z\rightarrow 1$.
However, the singularities will be compensated by the general expression of GDAs, since 
one can see $\Phi(z, \cos \theta, \hat{s}) \rightarrow 0$ 
if $ z\rightarrow 0$ or $ z\rightarrow 1$.
The dependence on energy scale $s$ is abbreviated in Eq.\,(\ref{eqn:gda-expression}),
in the limit of $s \rightarrow \infty$, only the terms with $n=1$  survive in the GDAs \cite{Diehl:2000uv},
\begin{align}
\Phi(z, \cos \theta, \hat{s})=18\, z(1-z) (2z-1) \left[\tilde{B}_{10}(\hat{s})
+\tilde{B}_{12}(\hat{s})  P_2(\cos \theta) \right],
\label{eqn:gda-exp-asm}
\end{align}
which is the so-called  asymptotic form of GDAs.
The first  and second terms correspond to the S-wave and
D-wave production of a meson pair, respectively.

\subsection{The $\gamma^*   \to \pi \eta  \gamma$ reaction : Search for exotic hybrid mesons and shear viscosity}

The  production of two different mesons, namely, $\gamma^*   \to \pi \eta  \gamma$ ,
where the $\pi \eta$ GDAs are involved, is very interesting from the point of view of exotic meson searches.
Indeed, in Ref.\,\cite{Anikin:2006du},  the authors proposed that $\gamma^{\ast} \gamma \to    \pi \eta$ can be used to search for 
the  resonances with $J^{PC}=1^{-+}$ by analyzing the $P$-wave of  $\pi \eta$, 
 and such resonances are candidates of the hybrid mesons  
 since the quantum numbers cannot exist for an ordinary meson in the quark model. Therefore, it is important to include
 the kinematical higher-twist corrections in the  helicity amplitudes of those reactions. The theoretical results of Eqs.\,(\ref{eqn:epho-cro}) and 
 (\ref{eqn:int-gdas-a}) can be used to describe $\gamma^*   \to M_1 M_2  \gamma$ except for  a few replacements,

\begin{align}
\beta_0 &  \rightarrow     \beta_0 = \sqrt{1-\frac{2(m_1^2+m_2^2)}{\hat{s}} +\frac{(m_1^2-m_2^2)^2}{\hat{s}^2} }, \nonumber \\
\Delta_T^2& \rightarrow    \Delta_T^2  =2(m_1^2+m_2^2)-(1-\zeta_0^2) \hat{s},  \nonumber \\
\zeta_0 & \rightarrow   \zeta_0=\beta_0 \cos{\theta}+\frac{(m_2^2-m_1^2)}{\hat{s}},
\label{eqn:rep}
\end{align} 
 where $m_1$ and $m_2$ denote the masses of $M_1$ and $M_2$, respectively\footnote{
As for $\gamma^*  \gamma  \to \pi \eta$, one can also use the theoretical expressions of Ref.\,\cite{Lorce:2022tiq} together with the first two replacements in Eq.\,(\ref{eqn:rep}), and the third one is slightly modified as 
$  
\zeta_0  \rightarrow   \zeta_0=-\beta_0 \cos{\theta}+\frac{(m_2^2-m_1^2)}{\hat{s}}.
$}.

 There are a few candidates for the isovector hybrid mesons, 
 for example $\pi_1(1400)$\cite{IHEP-Brussels:1988iqi,E862:2006cfp}, 
 $\pi_1(1600)$\cite{E852:1998mbq,COMPASS:2018uzl,JPAC:2018zyd,COMPASS:2021ogp} and
 $\pi_1(2015)$\cite{E852:2004gpn}, however, their existence is still controversial (see \cite{ParticleDataGroup:2022pth} and \cite{Chen:2022asf} for a recent review), and further confirmation is necessary.
In the near future, these  candidates  can be investigated in  $ \gamma^*   \to \pi \eta^{(\prime)} \gamma $  and  $ \gamma^*  \gamma  \to \pi \eta^{(\prime)}$, which are accessible at BESIII and Belle (Belle II).
Meanwhile, BESIII observed  a resonance called $\eta_1(1855)$  from the $P$-wave analysis of  $\eta \eta^{\prime}$  in the decay of
$ J/\Psi   \to \eta \eta^{\prime} \gamma $ very recently, which is a candidate of isoscalar hybrid mesons ($I^G(J^{PC})=(0^+)1^{-+}$)
\cite{BESIII:2022riz,BESIII:2022iwi}.
It will be promising to search for this resonance in $ \gamma^*  \to \eta \eta^{\prime} \gamma $, since one just needs to replace $J/\Psi$ with a timelike photon.

The  asymptotic GDAs are slightly modified by  the additional $P$-wave term in the production of two different scalar mesons
$M_1$ and $M_2$  \cite{Anikin:2006du},
\begin{align}
\Phi_q(z, \cos \theta, \hat{s})=30\, z(1-z) (2z-1) \left[\tilde{B}_{10}(\hat{s})+\tilde{B}_{11}(\hat{s}) P_1(\cos \theta)
+\tilde{B}_{12}(\hat{s})  P_2(\cos \theta) \right],
\label{eqn:gda-m1m2}
\end{align}
and the second term denotes the $P$-wave GDA, which is related to the production of exotic hybrid mesons.
We have $\Phi_u(z, \cos \theta, \hat{s})=\Phi_d(z, \cos \theta, \hat{s})$ and $\Phi_u(z, \cos \theta, \hat{s})=-\Phi_d(z, \cos \theta, \hat{s})$ for the total isospin $I=0$ and $I=1$ of the meson pairs, respectively.
The $M_1M_2$ GDAs can be also used to study the matrix element of the EMT,
\begin{align}
\langle M_2(p_2) M_1(p_1)  | \,T_q^{\mu \nu}\, | 0 \rangle \sim E_q(\hat{s}) P^{\mu} \Delta^{\nu},
\label{eqn:sh-vi}
\end{align}  
where $E_q(\hat{s})$ is  a new EMT form factor related to the shear viscosity;
 its sum over  quarks and gluons should be zero as a consequence of the conserved EMT, 
but however, $E_q(\hat{s})$ will exist for a single flavor $q$ on condition that there is $P$-wave GDA \cite{Teryaev:2022pke}. 
Thus, if one observes the candidates of the  hybrid mesons  in $ \gamma^*   \to \pi \eta^{(\prime)} \gamma $  and  $ \gamma^*  \gamma  \to \pi \eta^{(\prime)}$,  the existence of $E_q(\hat{s})$ will be proved by experiment.

\subsection{Numerical estimates of  the kinematical higher-twist contributions}
\begin{figure}[ht]
\centering
\includegraphics[width=0.85\textwidth]{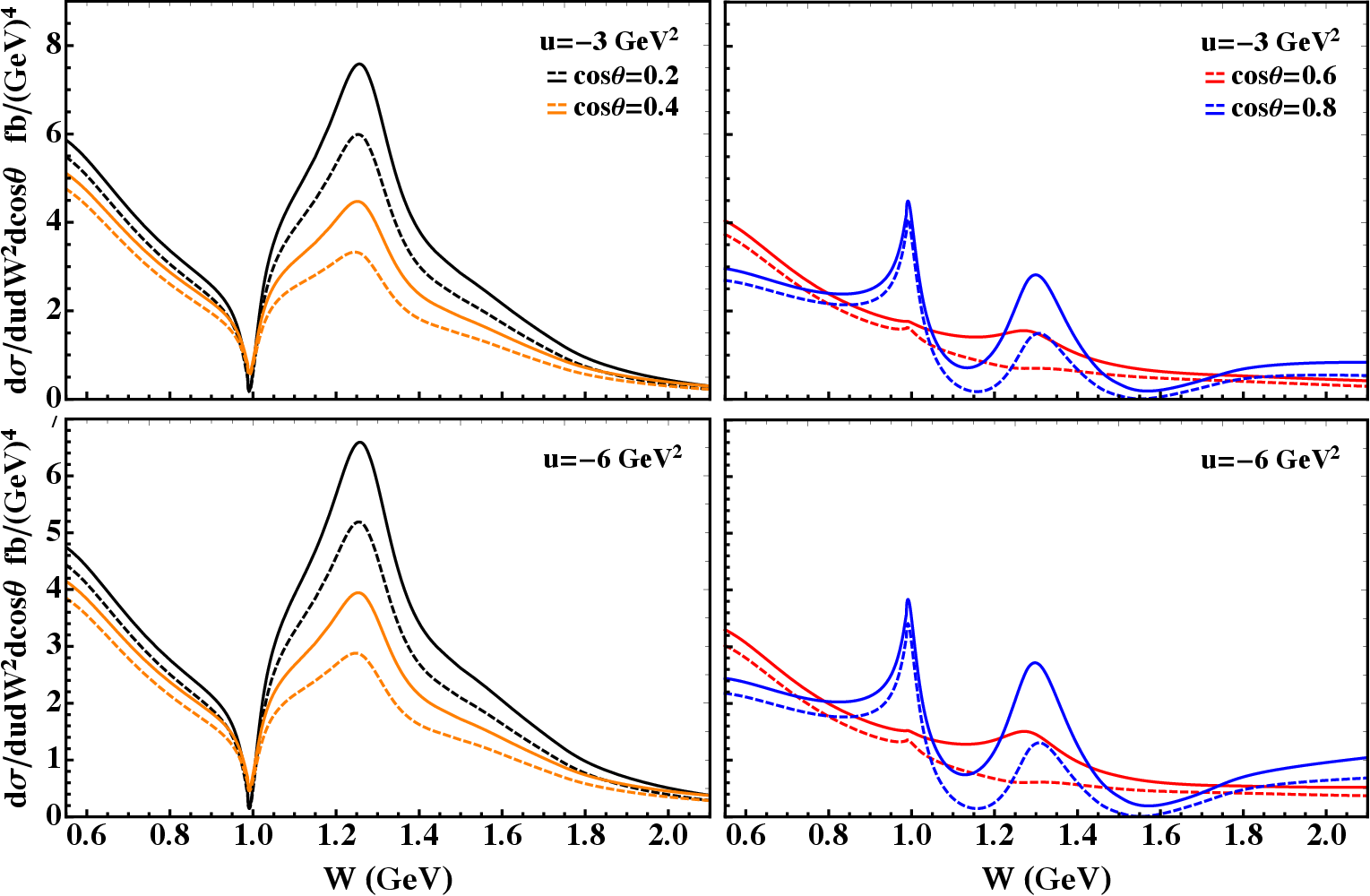}
\caption{ Differential cross section of  $e^- e^+  \to  \pi^0  \pi^0 \gamma$ is dependent on the invariant mass of pion pair 
$W=\sqrt{\hat{s}}$,
using the   $\pi \pi$ GDA extracted from Belle measurements \cite{Kumano:2017lhr}. The dashed lines are the twist-2 cross
sections, and the solid lines include the kinematical higher-twist contributions. }
\label{fig:num}
\end{figure}
In principle, the process  $\gamma^* \to M_1 M_2  \gamma$ can be measured by  BESIII and Belle (Belle II) in  $e^+ e^-$ collisions. The center-of-mass energy  is $\sqrt{s}=3-5$ GeV at BESIII, while it is $\sqrt{s}=8-10$ GeV
at Belle (Belle II). 
It should be much easier to measure this process at BESIII due to the  
 larger cross section which can be seen from Eq.\,(\ref{eqn:epho-cro}). As a consequence, we shall use the kinematics  of BESIII   in the numerical  estimate of the kinematical higher-twist contributions,
 and  the differential cross section of  Eq.\,(\ref{eqn:epho-cro}) is used by integrating over $\varphi$,
\begin{align}
\frac{d \sigma}{du\, dW^2\, d(\cos \theta)}=&
\frac{\alpha_{\text{em}}^3 \beta_0}{8s^3 } \,\frac{1}{1+\epsilon}   \,\Big[ |A_{++}|^2+ |A_{-+}|^2+2\epsilon\, |A_{0+}|^2  \Big],
\label{eqn:cro2}
\end{align} 
 where the helicity amplitudes are given by Eq.\,(\ref{eqn:int-gdas-a}) including the kinematical higher-twist contributions
  up to twist 4.

 We firstly calculate the cross section of  $e^- e^+ \to  \gamma^* \to \pi^0 \pi^0  \gamma$ 
 with the $\pi \pi$ GDA extracted from Belle measurements \cite{Kumano:2017lhr}.
In Fig.\,\ref{fig:num}, the twist-2 cross sections are depicted as  dashed lines, and the solid lines include the kinematical higher-twist contributions. 
The  kinematics is set  according to the BESIII experiment as $s=12 $ GeV$^2$ and $W \in (0.5, 2) $ GeV. 
The colors of the lines (black, orange, red, blue) represent  different values of $\cos \theta$ (0,2, 0.4, 0.6, 0.8), and $u$ is chosen as
$u= -3$ GeV$^2$ and $u= -6$ GeV$^2$. 
We can clearly see that the kinematical higher-twist corrections are always positive in the cross section, 
and this is different from the case of $e^-  \gamma \to e^- \pi^0 \pi^0 $ where the corrections can be positive or negative \cite{Lorce:2022tiq}. 
In the region  $W> 1$ GeV,  the kinematical higher-twist corrections turn out to be important and it is thus crucial to include them to extract in a valuable way GDAs from  the cross section of $e^- e^+ \to  \gamma^* \to \pi^0 \pi^0  \gamma$,
and then access both the timelike pion EMT form factors,
and  the spacelike ones,  obtained from the timelike ones by using dispersion relations requiring reliable information at $W> 1$ GeV  so as to make the integrals convergent.      
\begin{figure}[htp]
\centering
\includegraphics[width=0.85\textwidth]{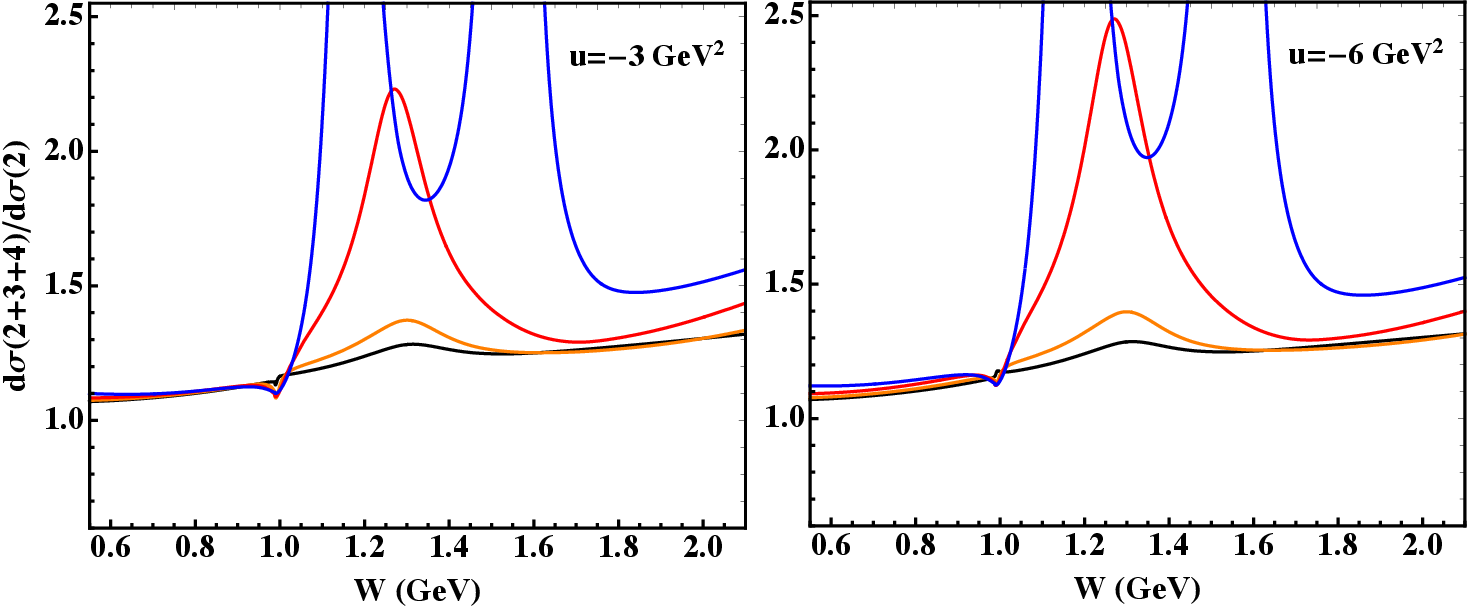}
\caption{ Ratio $d\sigma(2+3+4)/d\sigma(2)$ with the $\pi \pi$ GDA  extracted from Belle measurements, same conventions as in Fig.\,\ref{fig:num}.}
\label{fig:num2}
\end{figure}

The ratio $d\sigma(2+3+4)/d\sigma(2)$ is also shown in Fig.\,\ref{fig:num2}, where $d\sigma(i)$ ($i=2,3,4$) denotes the  twist-$i$ contribution to the cross section. The colors of the lines indicate different values of $\cos \theta$ as in Fig.\,\ref{fig:num}.
We can see that the kinematical higher-twist contributions have a significant impact on the cross section when $W> 1$ GeV.
The ratios just slightly change from $u=-3$ GeV$^2$ to $u=-6$ GeV$^2$, and this is because 
the ratios are dependent on $u$ through 
the parameter $\epsilon$,   namely, only the contribution from the amplitude $A_{0+}$ is affected in the ratios as one changes $u$. 
The peaks  
around $W \sim 1.1$ GeV and $W \sim 1.5$ GeV with $\cos \theta=0.8$ in Fig.\,\ref{fig:num2} 
are due to the fact that the twist-2 cross sections are quite tiny in this region as indicated by Fig.\,\ref{fig:num};
however, the extracted GDA used in this estimate  may not be accurate enough due to the large uncertainties of Belle measurements,
and these peaks in the ratio may thus not reflect  real physics.

\begin{figure}[htp]
\centering
\includegraphics[width=0.85\textwidth]{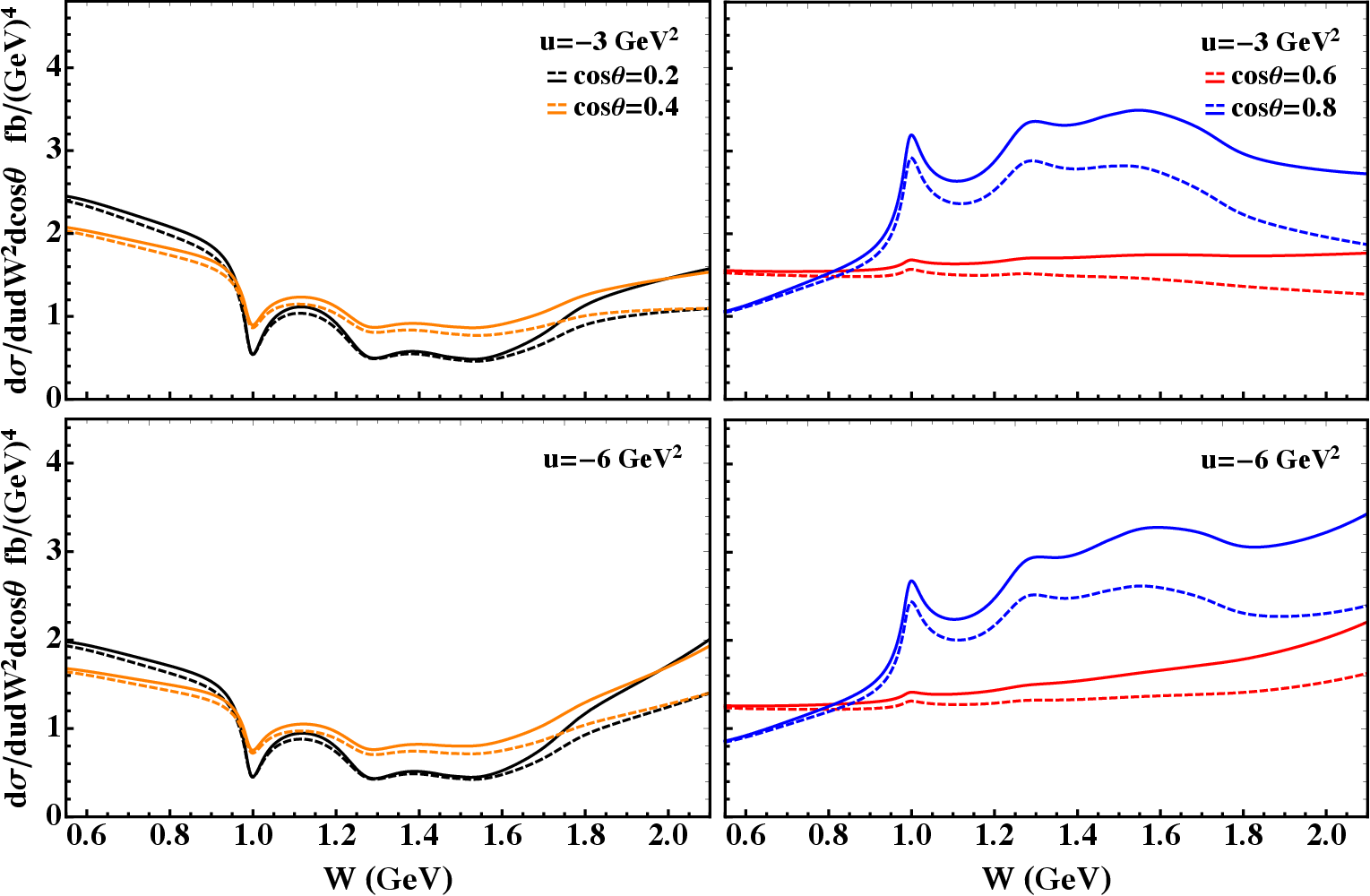}
\caption{ Differential cross section for  $e^-e^+  \to \gamma \to \pi^0  \pi^0 \gamma$ with the asymptotic $\pi \pi$ GDA described in the text, same conventions as in Fig.\,\ref{fig:num}.}
\label{fig:num3}
\end{figure}

For comparison, we also present our results in Fig.\,\ref{fig:num3} when we employ the asymptotic pion GDA  in the analysis of $e^-e^+  \to \gamma^{\ast} \to \pi^0  \pi^0 \gamma$.
In Ref.\,\cite{Diehl:1998dk}, the asymptotic GDA  was given when the energy scale $s \rightarrow \infty$,
\begin{align}
\Phi(z, \cos \theta, \hat{s})=20\, z(1-z)(2z-1) R_{\pi} \left[\frac{-3+\beta_0^2}{2}\, e^{i \delta_0} +
\beta_0^2 e^{i \delta_2}  P_2(\cos \theta)\right],
\label{eqn:gda-asy}
\end{align}
where $ R_{\pi}=0.5$ is the  momentum fraction carried by quarks in the pion meson.
$\delta_0$ is $\pi \pi$  the elastic scattering phase shift for S wave, and $\delta_2$ is the one for D wave \cite{Bydzovsky:2016vdx, Bydzovsky:2014cda, Surovtsev:2010cjf}.
The asymptotic pion GDA is indeed very different with the one extracted from Belle measurements, for example there is no contribution of resonance $f_2$ in the asymptotic GDA. However, the main purpose of this work is not to estimate  cross sections accurately, but to see whether the kinematical corrections are sizeable or not.
In Fig.\,\ref{fig:num3}, 
 $u$ is chosen as $u=$-3 GeV$^2$ and $u=$-6 GeV$^2$ together with  0.5 GeV $  \leq  W \leq $ 2.1 GeV and $s=12$ GeV$^2$.
The dashed lines represent  the twist-2 cross sections, and the solid ones include kinematical higher-twist contributions.
 Black lines  denote $\cos \theta=0.2$ and orange lines correspond to $\cos \theta=0.4$, while $\cos \theta=0.6$ and  $\cos \theta=0.8$ are depicted as red and blue, respectively.
 We can clearly see that the kinematical higher-twist corrections become more and more important as one increases $W=\sqrt{\hat{s}}$ in this figure, which is evident since  the corrections are expected to be proportional to 
$\hat{s}/s$.
As for  the case of the extracted GDA  from Belle measurements, it is thus necessary to include the kinematical higher-twist corrections to describe the cross section in the region of $W > 1$ GeV. Both  GDAs indicate a similar magnitude of the cross section for  $e^-e^+  \to \gamma^{\ast} \to \pi^0  \pi^0 \gamma$.
 \begin{figure}[htp]
\centering
\includegraphics[width=0.85\textwidth]{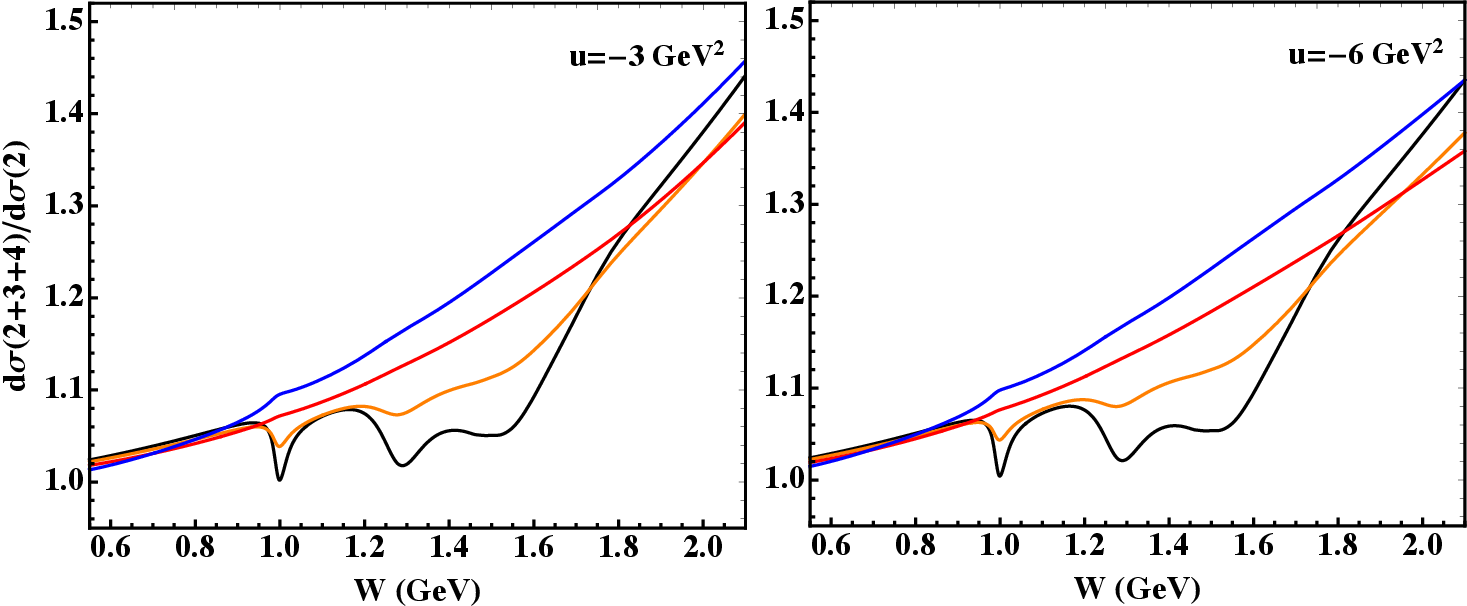}
\caption{Ratio $d\sigma(2+3+4)/d\sigma(2)$ with the  asymptotic $\pi \pi$ GDA described in the text, same conventions as in Fig.\,\ref{fig:num}.}
\label{fig:num4}
\end{figure}

In Fig.\,\ref{fig:num4}, the ratio of  $d\sigma(2+3+4)/d\sigma(2)$ is also presented  so as to
see  the proportion of the kinematical higher-twist corrections in the cross section clearly.
 $u$ is set as  $u=$-3 GeV$^2$ and $u=$-6 GeV$^2$ for the left panel and right panel, respectively,
and the colors of lines indicate the values of
$\cos \theta$ as in Fig.\,\ref{fig:num3}.
The ratios increase rapidly as $W$ goes up, and the kinematical higher-twist corrections
 account for more than $40\%$ of the cross section around $W\sim2 $ GeV, which proves that they need to be included in any reliable GDA extraction from the cross section.
\begin{figure}[htp]
\centering
\includegraphics[width=0.85\textwidth]{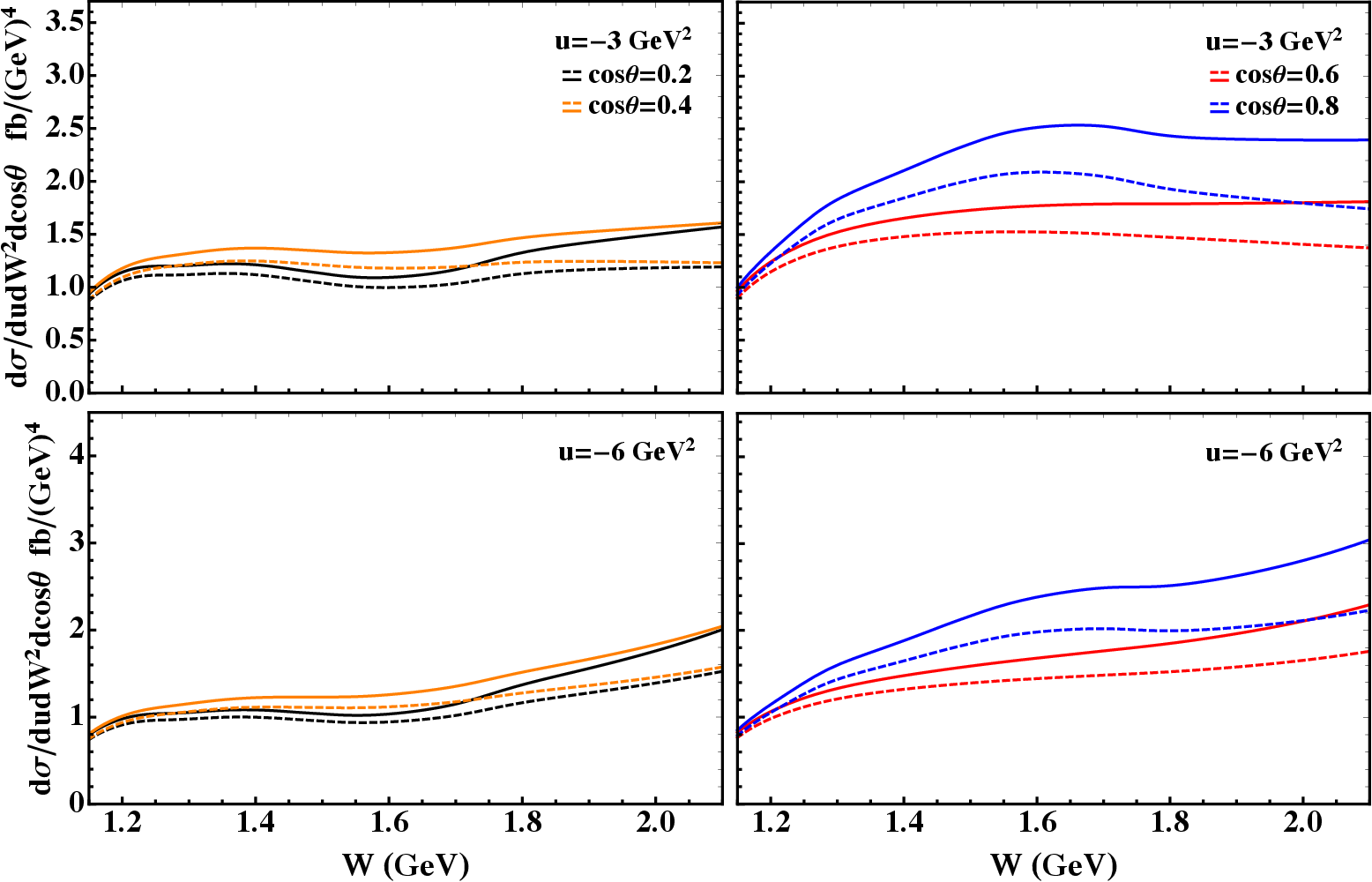}
\caption{ Differential cross section for  $e^-e^+  \to \gamma^{\ast} \to \eta  \eta \gamma$ with the model $\eta \eta$ GDA described in the text, same conventions as in Fig.\,\ref{fig:num}.}
\label{fig:num5}
\end{figure}

\begin{figure}[htp]
\centering
\includegraphics[width=0.85\textwidth]{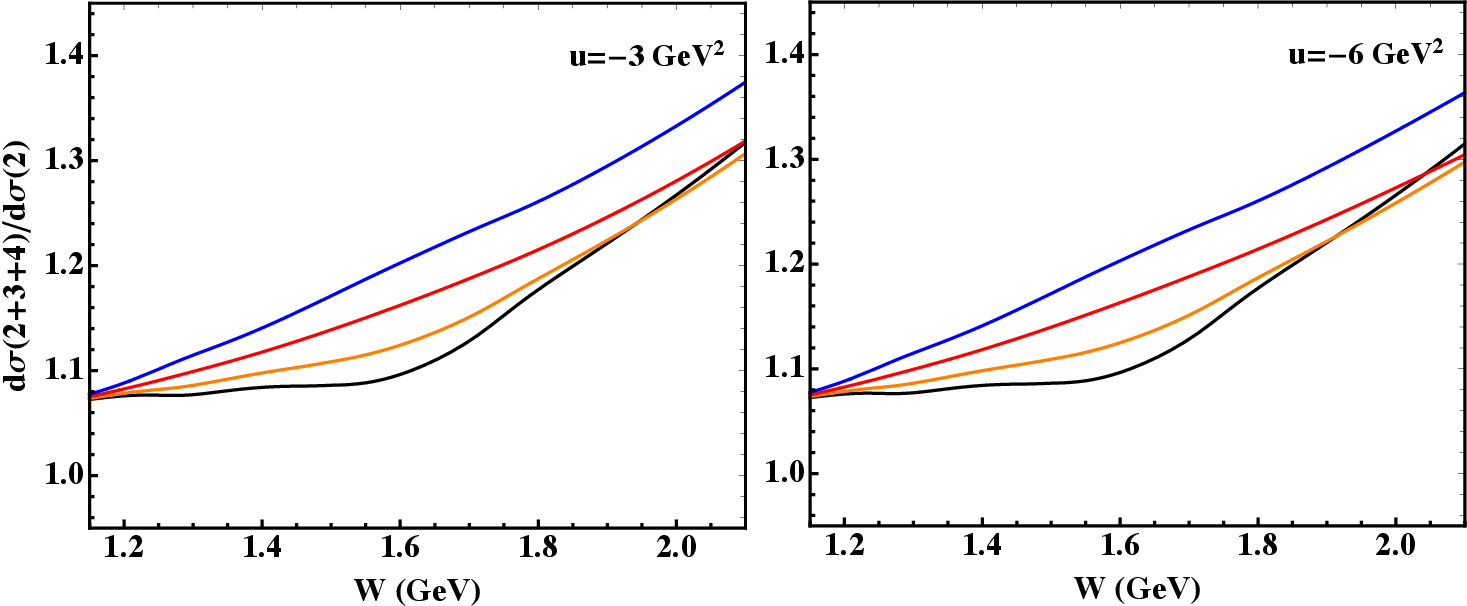}
\caption{Ratio $d\sigma(2+3+4)/d\sigma(2)$ with the model  $\eta \eta$ GDA described in the text, same conventions as in Fig.\,\ref{fig:num}.}
\label{fig:num6}
\end{figure}

There are two types of corrections $m^2/s$ and $\hat{s}/s$ in the kinematical  corrections, and only the latter contributes to
the cross section of $e^-e^+  \to \gamma^{\ast} \to \pi^0  \pi^0 \gamma$ due to the small mass of pion meson in comparison with 
the value of $s$.
In order to see the impact of the target mass correction $m^2/s$, we now consider the production of a pair of slightly heavier mesons, namely $\eta \eta$.
However, very little information is known about their GDAs at the current stage, and  we thus estimate the kinematical  higher-twist corrections for $e^-e^+  \to \gamma^{\ast} \to \eta  \eta \gamma$ by using a simple model GDA 
identical to  the asymptotic $\pi \pi$ GDA except that the  $\eta$ mass is used. The center-of-mass energy of 
$e^-e^+$ is again chosen as $s=12$ GeV$^2$.
In Fig.\,\ref{fig:num5}, the cross sections are shown with the range of 1.2 GeV $  \leq  W \leq $ 2.1 GeV.
The colors of the lines denote different values of $\cos \theta$ as indicated on the different panels of the figure.
The dashed lines represent  the twist-2 cross sections, and the solid ones include kinematical higher-twist contributions.
The gaps between the dashed lines and the solid ones increase  along with $W$ as  expected, and one thus cannot neglect the kinematical higher-twist contributions in the cross section.
The kinematical higher-twist contributions are always positive in the cross section.
We present the ratios of  $d\sigma(2+3+4)/d\sigma(2)$  in Fig.\,\ref{fig:num6}.
The kinematical higher-twist contributions 
account for  less than $40\%$ of the cross section around $W\sim2 $ GeV.
Compared with the results in Fig.\,\ref{fig:num4}, the ratios decrease if one replace the mass of the $\pi$ meson by the one of the
$\eta$ meson, keeping  the  asymptotic $\pi \pi$ GDA as a model for the $\eta \eta$ GDA ; indeed, the target mass corrections  are negative and thus diminish the positive corrections of order $\hat{s}/s$ in the cross section.
 
We do not plot the kinematical higher twist contributions for the $\pi \eta$ production case, since they depend much on the unknown $\pi \eta$ GDAs. One can estimate that they are somewhat in between the relative contributions in the $\pi \pi$ and $\eta \eta$ 
cases displayed in Fig. \ref{fig:num4} and Fig. \ref{fig:num6}.

\section{Summary}
\label{summary}

GDAs  can be studied  in both $\gamma^*  \gamma \to M_1 M_2$ and $\gamma^*   \to M_1 M_2 \gamma$.
The former process has been measured by Belle for $\pi \pi$ \cite{Belle:2015oin} and $KK$ \cite{Belle:2017xsz} with  large uncertainties; in the near future we can expect more precise measurements from Belle II due to the much higher luminosity.
It will be more advantageous to measure the latter process at BESIII,
since its cross section will be suppressed by the larger center-of-mass energy of electron-positron pair at Belle (Belle II).
In this case, the measurements of $\gamma^*  \gamma \to M_1 M_2$
 at Belle and Belle II can be cross checked by the ones of 
 $\gamma^*   \to M_1 M_2 \gamma$ 
 at BESIII due to the similar kinematics,
and the GDAs can be extracted by combining the measurements of the two pocesses.
Besides, since GDAs are probed by a spacelike photon and a timelike photon in $\gamma^*  \gamma \to M_1 M_2$ and $\gamma^*   \to M_1 M_2 \gamma$, respectively, we can also check the university of GDAs.

In this work we calculate the kinematical higher-twist corrections for $\gamma^*  \to M_1 M_2 \gamma$ up to twist 4,
and three helicity amplitudes are expressed in term of the twist-2 GDA.
We calculated the cross section with and without the kinematic higher twist contributions in terms of the leading twist GDAs.
We adopt two types of GDAs to estimate the kinematical higher-twist contributions for $\gamma^*  \to \pi^0 \pi^0 \gamma$ numerically.
In the calculation, the center-of-mass energy of the electron-positron pair is chosen as $s=12$ GeV$^2$, which is typical for BESIII.
All the numerical results indicate that the kinematical higher-twist corrections have a significant impact on the cross section of 
 $\gamma^*  \to M_1 M_2 \gamma$ as in the case of  $\gamma^*  \gamma \to M_1 M_2$ \cite{Lorce:2022tiq, Lorce:2022cze}. However, the corrections
are always positive in the cross section of the former, and this is different from the latter process where kinematical higher-twist corrections can go both ways. A model $\eta \eta$ GDA is used to see the impact of the target mass corrections of $\mathcal O(m^2/s)$,
and the kinematical higher-twist corrections account for about $20\%$ of the cross section  in the region of 1.2 GeV $\le W \le $2.1 GeV on the average, which are not negligible. 
As a consequence, it is important to use the accurate description of the cross section with the inclusion of kinematical contributions when one tries to extract GDAs from experimental measurements. 
The extracted GDAs can be used to study the  EMT form factors  of hadrons, 
which are important physical quantities to investigate 
mass, pressure and shear force distributions of hadrons.

The present study was performed at lowest order in the strong coupling, but it would be interesting to include higher order corrections which are known - at leading twist - to be very sensitive to the timelike vs spacelike nature of the probe \cite{Mueller:2012sma}.

\section*{Acknowledgments}
We acknowledge useful discussions with C\'edric Lorc\'e, Wen-Cheng Yan and Ya-Teng Zhang. Qin-Tao Song was supported by the National Natural Science Foundation 
of China under Grant Number 12005191.

\appendix
\section{Helicity amplitudes in terms of DDs}
\label{app-b}
The double distributions (DDs) of a scalar meson are defined by \cite{Teryaev:2001qm} 
\begin{align}
\langle \bar{M}(p_2) M(p_1)  | \, \bar{q}(z_1 n) \slashed{n}q(z_2 n)  \,  | 0 \rangle
=  \int d\beta\, d\alpha   \left[f_q(\beta, \alpha) \,\Delta\cdot n- g_q(\beta, \alpha)\, 2P\cdot n   \right] 
e^{-i  l_{z_1 z_2}\cdot n},
\label{eqn:dds}
\end{align}
where the support region of $f_q$ and $g_q$ is given by the rhombus $|\alpha|+|\beta|\leq 1$, and the momentum $l_{z_1 z_2}$ is written as
\begin{align}
l_{z_1 z_2}=(z_2-z_1) \left[ \beta\, \frac{\Delta}{2} -(\alpha+1) P  \right] -2z_1 P.
\label{eqn:lz1z2}
\end{align}
If one combines Eqs.\,(\ref{eqn:gda}) with (\ref{eqn:dds}),
the GDA can be expressed in terms of DDs,
\begin{align}
\Phi_q(z,\zeta_0, s)=2\int d\beta \, d\alpha\,  \delta(y+\alpha-\beta\zeta_0 ) 
\left[f_q(\beta, \alpha)\, \zeta_0 - g_q(\beta, \alpha) \right], 
\label{eqn:dd-gda}
\end{align}
where $y=2z-1$.
Assuming that the DDs vanish at the boundaries, Eq.\,(\ref{eqn:dds}) can be expressed as \cite{Lorce:2022tiq}
\begin{equation}
\langle \bar{M}(p_2) M(p_1)  | \, \bar{q}(z_1 n) \slashed{n}q(z_2 n)  \, | 0 \rangle
=  \frac{2i}{z_{12}}  \int d\beta  \,d\alpha\, 
 \phi_q(\beta, \alpha) \,
 e^{-i l_{z_1 z_2}\cdot n}, \label{eqn:dds-mod}
\end{equation}
where the notation $z_{12}=z_1- z_2$ is used, and $\phi_q(\beta, \alpha)$  is defined by
\begin{align}
\phi_q(\beta, \alpha) =  \partial_{\beta}f_q(\beta, \alpha) + \partial_{\alpha}g_q(\beta, \alpha).
\label{eqn:dds-new}
\end{align}
Due to even charge conjugation of the meson pair, we can have the symmetry $\phi_q(\beta, \alpha) = \phi_q(\beta,-\alpha) =\phi_q(-\beta, -\alpha)$, which is used to simplify the calculation of the amplitudes.

The leading-twist operator $\mathcal{O}_{++}^{t=2}(z_1 , z_2 )$ appears in  the kinematical contributions of Eq.\,(\ref{eqn:kine-t4}), where the separation $x$ is not necessarily lightlike. However, GDAs and DDs are defined by the matrix element of $O_{++}(z_1 n, z_2 n)$ with the lightlike separation $n$ as shown in Eq.\,(\ref{eqn:dds}),
\begin{align}
O_{++}(z_1 n, z_2 n)= \sum_q e_q^2\,\bar{q}(z_1 n) \slashed{n} q(z_2 n).
\label{eqn:optb2}
\end{align}
The matrix element of $\mathcal{O}_{++}^{t=2}(z_1 , z_2 )$  is related to the one of 
$O_{++}(z_1 n, z_2 n)$ by using the leading-twist projector $\Pi(x, n)$ \cite{Braun:2011dg,Braun:2011zr,Braun:2011th},
\begin{align}
\langle \bar{M}(p_2) M(p_1)  | \, \mathcal{O}_{++}^{t=2}(z_1 , z_2 )  \, | 0 \rangle
= \Pi(x, n) \langle \bar{M}(p_2) M(p_1)  | \, O_{++}(z_1 n, z_2 n)  \,  | 0 \rangle.
\label{eqn:projector}
\end{align}
If one combines Eqs.\,(\ref{eqn:projector}) and (\ref{eqn:dds-mod}),  the matrix element of $\mathcal{O}_{++}^{t=2}(z_1 , z_2 )$ can be obtained \cite{Lorce:2022tiq},
\begin{align}
\langle \bar{M}(p_2) M(p_1)  | \, \mathcal{O}_{++}^{t=2}(z_1 , z_2 )  \,  | 0 \rangle
= \chi \, \frac{2i}{z_{12}}  \int d\beta\,  d\alpha 
\,\phi(\beta, \alpha) \left[ e^{-i l_{z_1 z_2}\cdot x}   + \frac{x^2 l_{z_1z_2}^2}{4} \int_0^1  dv\, v  \,
e^{-i v l_{z_1 z_2}\cdot x}  \right ],
\label{eqn:dds-t4}
\end{align}
where $\phi=\phi_u+\phi_d$  and  $\chi= 5e^2/18$ for an isosinglet meson pair.
Furthermore, the matrix elements of   $\mathcal{O}_1$ and $\mathcal{O}_2$ can be given  by 
\begin{align}
\langle \bar{M}(p_2) M(p_1)  | \, \mathcal{O}_1(z_1, z_2) \,  | 0 \rangle
&= -\chi \, \frac{2i}{z_{12}}\, \hat{s}  \int d\beta\,d\alpha   \,
\phi(\beta, \alpha) \, e^{-i l_{z_1 z_2}\cdot x}, \nonumber \\
\langle \bar{M}(p_2) M(p_1)  | \, \mathcal{O}_2(z_1, z_2) \,  | 0 \rangle
&= \chi\,  \frac{2i}{z_{12}}  \int d\beta\,d\alpha   \,
\phi(\beta, \alpha) \left[ 2P\cdot l_{z_1 z_2} \,  e^{-i l_{z_1 z_2}\cdot x}   
+i P \cdot x \,l_{z_1z_2}^2 \int_0^1  dv\, v  \,
e^{-i v l_{z_1 z_2}\cdot x}  \right ],
\label{eqn:o1o2-t4}
\end{align}
where
\begin{align}
\mathcal{O}_1(z_1, z_2)&= \left[i \mathbf{P}^{\mu}, \, \left[i \mathbf{P}_{\mu}, \, \mathcal{O}_{++}^{t=2}(z_1 , z_2 )\right] \right], \nonumber \\
\mathcal{O}_2(z_1, z_2)&= 
\left[i \mathbf{P}^{\mu}, \, \frac{\partial}{\partial x^{\mu}}\mathcal{O}_{++}^{t=2}(z_1 , z_2 ) \right].
\label{eqn:total-deri}
\end{align}

The helicity amplitudes are expressed in terms of matrix elements of operators, 
which are shown in Eqs.\,(\ref{eqn:dds-t4}) and (\ref{eqn:o1o2-t4}).
One obtains after a lengthy calculation
\begin{align}
A_{0-}&=-2  \chi \,  \frac{ \Delta \cdot \epsilon_{+} }{\sqrt{s}}
\int d\beta \,  d\alpha\, \phi(\beta, \alpha)\, \beta\,
\frac{\ln(F)  }{F-1}, \nonumber \\
A_{+-}&
= \chi\,  \frac{(\Delta \cdot \epsilon_{+})^2}{2 n\cdot \tilde{n}}  \int d\beta  \,d\alpha \,\phi(\beta, \alpha)\,    \beta^2\,
\partial_F  \left[ \frac{1-2F}{F-1} \,\ln(F)   \right],\nonumber \\
A_{++}&=\chi   \int d\beta\,   d\alpha\,  \phi(\beta, \alpha)
\left\{ 2 \ln(F)  -  \left[ \frac{\hat{s}}{n \cdot\tilde{n}}\,(F-\alpha) + \frac{\beta^2 \Delta_T^2  }{4n\cdot \tilde{n}} \,\partial_F    \right]   \frac{1}{F-1}   \left[ \frac{\ln(F)}{2}-\text{Li}_2(1)+ \text{Li}_2(F) \right]    \right\},
\label{eqn:a11-all-final}
\end{align}
where  $F$ is defined as
\begin{align}
F(\alpha, \beta) = \frac{\alpha -\beta \zeta_0 +1}{2}.
\label{eqn:F}
\end{align}
The helicity amplitudes can be presented in terms of the GDA using  \cite{Lorce:2022tiq}
\begin{align}
\frac{\partial \Phi_q(z,\zeta_0, s)}{\partial z} =4\int d\beta\, d\alpha\, \delta((2z-1)+\alpha-\beta \zeta_0)\, \phi_q(\beta, \alpha).
\label{eqn:dds-gda}
\end{align}




\begin{thebibliography}{00}




\bibitem{Muller:1994ses}
D.~M\"uller, D.~Robaschik, B.~Geyer, F.~M.~Dittes and J.~Ho\v{r}ej\v{s}i,
Fortsch. Phys. \textbf{42} (1994), 101-141.


\bibitem{Diehl:1998dk}
M.~Diehl, T.~Gousset, B.~Pire and O.~Teryaev,
Phys. Rev. Lett. \textbf{81} (1998), 1782-1785.

\bibitem{Diehl:2000uv}
M.~Diehl, T.~Gousset and B.~Pire,
Phys. Rev. D \textbf{62} (2000), 073014.


\bibitem{Polyakov:1998ze}
M.~V.~Polyakov,
Nucl. Phys. B \textbf{555} (1999), 231.


\bibitem{Pire:2002ut}
B.~Pire and L.~Szymanowski,
Phys. Lett. B \textbf{556} (2003), 129-134.





\bibitem{Diehl:2003ny}
M.~Diehl,
Phys. Rept. \textbf{388} (2003), 41-277.


\bibitem{Belitsky:2005qn}
A.~V.~Belitsky and A.~V.~Radyushkin,
Phys. Rept. \textbf{418} (2005), 1-387.


\bibitem{Boffi:2007yc}
S.~Boffi and B.~Pasquini,
Riv. Nuovo Cim. \textbf{30} (2007) no.9, 387-448.






\bibitem{Goeke:2001tz}
K.~Goeke, M.~V.~Polyakov and M.~Vanderhaeghen,
Prog. Part. Nucl. Phys. \textbf{47} (2001), 401-515.








\bibitem{Belle:2015oin}
M.~Masuda \textit{et al.} [Belle],
Phys. Rev. D \textbf{93} (2016) no.3, 032003.

\bibitem{Belle:2017xsz}
M.~Masuda \textit{et al.} [Belle],
Phys. Rev. D \textbf{97} (2018) no.5, 052003.


\bibitem{Kumano:2017lhr}
S.~Kumano, Qin-Tao~Song and O.~V.~Teryaev,
Phys. Rev. D \textbf{97} (2018), 014020.



\bibitem{Lu:2006ut}
Z.~Lu and I.~Schmidt,
Phys. Rev. D \textbf{73} (2006), 094021
[erratum: Phys. Rev. D \textbf{75} (2007), 099902].


\bibitem{BaBar:2015onb}
J.~P.~Lees \textit{et al.} [BaBar],
Phys. Rev. D \textbf{92} (2015) no.7, 072015.









\bibitem{Chen:2002th}
C.~H.~Chen and H.~N.~Li,
Phys. Lett. B \textbf{561} (2003), 258-265.


\bibitem{Wang:2015uea}
W.~F.~Wang, H.~N.~Li, W.~Wang and C.~D.~L\"u,
Phys. Rev. D \textbf{91} (2015) no.9, 094024.


\bibitem{Li:2016tpn}
Y.~Li, A.~J.~Ma, W.~F.~Wang and Z.~J.~Xiao,
Phys. Rev. D \textbf{95} (2017) no.5, 056008.



\bibitem{Jia:2021uhi}
M.~K.~Jia, C.~Q.~Zhang, J.~M.~Li and Z.~Rui,
Phys. Rev. D \textbf{104} (2021) no.7, 073001.



\bibitem{Burkardt:2000za}
M.~Burkardt,
Phys. Rev. D \textbf{62} (2000), 071503
[erratum: Phys. Rev. D \textbf{66} (2002), 119903].

\bibitem{Ralston:2001xs}
J.~P.~Ralston and B.~Pire,
Phys. Rev. D \textbf{66} (2002), 111501.


\bibitem{Diehl:2002he}
M.~Diehl,
Eur. Phys. J. C \textbf{25} (2002), 223-232
[erratum: Eur. Phys. J. C \textbf{31} (2003), 277-278].




\bibitem{Ji:1996ek}
X.~D.~Ji,
Phys. Rev. Lett. \textbf{78} (1997), 610-613.


\bibitem{Ji:1996nm}
X.~D.~Ji,
Phys. Rev. D \textbf{55} (1997), 7114-7125.







\bibitem{Polyakov:2002yz}
M.~V.~Polyakov,
Phys. Lett. B \textbf{555} (2003), 57-62.


\bibitem{Goeke:2007fp}
K.~Goeke, J.~Grabis, J.~Ossmann, M.~V.~Polyakov, P.~Schweitzer, A.~Silva and D.~Urbano,
Phys. Rev. D \textbf{75} (2007), 094021.


\bibitem{Mai:2012cx}
M.~Mai and P.~Schweitzer,
Phys. Rev. D \textbf{86} (2012), 096002.


\bibitem{Polyakov:2018zvc}
M.~V.~Polyakov and P.~Schweitzer,
Int. J. Mod. Phys. A \textbf{33} (2018) no.26, 1830025.

\bibitem{Lorce:2018egm}
C.~Lorc\'e, H.~Moutarde and A.~P.~Trawi\'nski,
Eur. Phys. J. C \textbf{79} (2019) no.1, 89.

\bibitem{Burkert:2018bqq}
V.~D.~Burkert, L.~Elouadrhiri and F.~X.~Girod,
Nature \textbf{557} (2018) no.7705, 396-399.

\bibitem{Kumericki:2019ddg}
K.~Kumeri\v{c}ki,
Nature \textbf{570} (2019) no.7759, E1-E2.

\bibitem{Dutrieux:2021nlz}
H.~Dutrieux, C.~Lorc\'e, H.~Moutarde, P.~Sznajder, A.~Trawi\'nski and J.~Wagner,
Eur. Phys. J. C \textbf{81} (2021) no.4, 300.

\bibitem{Burkert:2023wzr}
V.~D.~Burkert, L.~Elouadrhiri, F.~X.~Girod, C.~Lorc\'e, P.~Schweitzer and P.~E.~Shanahan,
[arXiv:2303.08347 [hep-ph]].



\bibitem{Anikin:2006du}
I.~V.~Anikin, B.~Pire, L.~Szymanowski, O.~V.~Teryaev and S.~Wallon,
Eur. Phys. J. C \textbf{47} (2006), 71-79.


\bibitem{Teryaev:2022pke}
O.~Teryaev,
JPS Conf. Proc. \textbf{37}, 020406 (2022)
doi:10.7566/JPSCP.37.020406
[arXiv:2204.09742 [hep-ph]].



\bibitem{Braun:2011zr}
V.~M.~Braun and A.~N.~Manashov,
Phys. Rev. Lett. \textbf{107} (2011), 202001.


\bibitem{Braun:2011dg}
V.~M.~Braun and A.~N.~Manashov,
JHEP \textbf{01} (2012), 085.


\bibitem{Braun:2011th}
V.~M.~Braun and A.~N.~Manashov,
Prog. Part. Nucl. Phys. \textbf{67} (2012), 162-167.


\bibitem{Braun:2022qly}
V.~M.~Braun, Y.~Ji and A.~N.~Manashov,
JHEP \textbf{01} (2023), 078.

\bibitem{Nachtmann:1973mr}
O.~Nachtmann,
Nucl. Phys. B \textbf{63} (1973), 237-247.


\bibitem{Sato:2016tuz}
N.~Sato \textit{et al.} [Jefferson Lab Angular Momentum],
Phys. Rev. D \textbf{93} (2016) no.7, 074005.


\bibitem{Braun:2012bg}
V.~M.~Braun, A.~N.~Manashov and B.~Pirnay,
Phys. Rev. D \textbf{86} (2012), 014003.


\bibitem{Braun:2012hq}
V.~M.~Braun, A.~N.~Manashov and B.~Pirnay,
Phys. Rev. Lett. \textbf{109} (2012), 242001.

\bibitem{Braun:2014sta}
V.~M.~Braun, A.~N.~Manashov, D.~M\"uller and B.~M.~Pirnay,
Phys. Rev. D \textbf{89} (2014), 074022.


\bibitem{JeffersonLabHallA:2022pnx}
F.~Georges \textit{et al.} [Jefferson Lab Hall A],
Phys. Rev. Lett. \textbf{128} (2022), 252002.




\bibitem{Lorce:2022tiq}
C.~Lorc\'e, B.~Pire and Qin-Tao~Song,
Phys. Rev. D \textbf{106} (2022), 094030.

\bibitem{Lorce:2022cze}
C.~Lorc\'e, B.~Pire and Qin-Tao~Song,
[arXiv:2208.12532 [hep-ph]].

\bibitem{JeffersonLabHallA:2015dwe}
M.~Defurne \textit{et al.} [Jefferson Lab Hall A],
Phys. Rev. C \textbf{92} (2015), 055202.


\bibitem{Defurne:2017paw}
M.~Defurne \textit{et al.},
Nature Commun. \textbf{8} (2017) no.1, 1408.





\bibitem{Braun:2008ia}
V.~M.~Braun, A.~N.~Manashov and J.~Rohrwild,
Nucl. Phys. B \textbf{807} (2009), 89-137.



\bibitem{Braun:2009vc}
V.~M.~Braun, A.~N.~Manashov and J.~Rohrwild,
Nucl. Phys. B \textbf{826} (2010), 235-293.

\bibitem{Teryaev:2001qm}
O.~V.~Teryaev,
Phys. Lett. B \textbf{510} (2001), 125-132.



\bibitem{IHEP-Brussels:1988iqi}
D.~Alde \textit{et al.} [IHEP-Brussels-Los Alamos-Annecy(LAPP)],
Phys. Lett. B \textbf{205} (1988), 397.



\bibitem{E862:2006cfp}
G.~S.~Adams \textit{et al.} [E862],
Phys. Lett. B \textbf{657} (2007), 27-31.
[arXiv:hep-ex/0612062 [hep-ex]].






\bibitem{E852:1998mbq}
G.~S.~Adams \textit{et al.} [E852],
Phys. Rev. Lett. \textbf{81} (1998), 5760-5763.




\bibitem{COMPASS:2018uzl}
M.~Aghasyan \textit{et al.} [COMPASS],
Phys. Rev. D \textbf{98} (2018) no.9, 092003



\bibitem{JPAC:2018zyd}
A.~Rodas \textit{et al.} [JPAC],
Phys. Rev. Lett. \textbf{122} (2019), 042002.



\bibitem{COMPASS:2021ogp}
M.~G.~Alexeev \textit{et al.} [COMPASS],
Phys. Rev. D \textbf{105} (2022), 012005.




\bibitem{E852:2004gpn}
J.~Kuhn \textit{et al.} [E852],
Phys. Lett. B \textbf{595} (2004), 109-117.





\bibitem{ParticleDataGroup:2022pth}
R.~L.~Workman \textit{et al.} [Particle Data Group],
PTEP \textbf{2022} (2022), 083C01.

\bibitem{Chen:2022asf}
H.~X.~Chen, W.~Chen, X.~Liu, Y.~R.~Liu and S.~L.~Zhu,
Rept. Prog. Phys. \textbf{86} (2023) no.2, 026201.




\bibitem{BESIII:2022riz}
M.~Ablikim \textit{et al.} [BESIII],
Phys. Rev. Lett. \textbf{129} (2022) no.19, 192002
[erratum: Phys. Rev. Lett. \textbf{130} (2023) no.15, 159901].

\bibitem{BESIII:2022iwi}
M.~Ablikim \textit{et al.} [BESIII],
Phys. Rev. D \textbf{106} (2022) no.7, 072012
[erratum: Phys. Rev. D \textbf{107} (2023) no.7, 079901].










\bibitem{Bydzovsky:2016vdx}
P.~Byd\v{z}ovsk\'y, R.~Kami\'nski and V.~Nazari,
Phys. Rev. D \textbf{94} (2016) 11, 116013.




\bibitem{Bydzovsky:2014cda}
P.~Byd\v{z}ovsk\'y, R.~Kami\'nski and V.~Nazari,
Phys. Rev. D \textbf{90} (2014)11, 116005.





\bibitem{Surovtsev:2010cjf}
Y.~S.~Surovtsev, P.~Bydzovsky, R.~Kaminski and M.~Nagy,
Phys. Rev. D \textbf{81} (2010), 016001.

\bibitem{Mueller:2012sma}
D.~Mueller, B.~Pire, L.~Szymanowski and J.~Wagner,
Phys. Rev. D \textbf{86} (2012), 031502. 





















\end{thebibliography}
\end{document}